\documentclass[twocolumn,aps,prc,superscriptaddress]{revtex4-2}
\usepackage{amsmath,amssymb,amsfonts}
\usepackage{bm}
\usepackage{graphicx}
\usepackage{mathrsfs}
\usepackage{multirow}
\usepackage{graphicx}
\usepackage{booktabs, ctable}
\usepackage{xcolor, color, framed}
\usepackage{subcaption}
\usepackage[colorlinks, linkcolor=red,anchorcolor=green,citecolor=blue]{hyperref}

\begin{document}

\newcommand{\vk}{{\vec k}}
\newcommand{\vK}{{\vec K}}
\newcommand{\vb}{{\vec b}}
\newcommand{{\vp}}{{\vec p}}
\newcommand{{\vq}}{{\vec q}}
\newcommand{\vQ}{{\vec Q}}
\newcommand{\vx}{{\vec x}}
\newcommand{\beq}{\begin{equation}}
\newcommand{\eeq}{\end{equation}}
\newcommand{\half}{{\textstyle \frac{1}{2}}}
\newcommand{\gton}{\stackrel{>}{\sim}}
\newcommand{\lton}{\mathrel{\lower.9ex \hbox{$\stackrel{\displaystyle<}{\sim}$}}}
\newcommand{\ee}{\end{equation}}
\newcommand{\ben}{\begin{enumerate}}
\newcommand{\een}{\end{enumerate}}
\newcommand{\bit}{\begin{itemize}}
\newcommand{\eit}{\end{itemize}}
\newcommand{\bc}{\begin{center}}
\newcommand{\ec}{\end{center}}
\newcommand{\bea}{\begin{eqnarray}}
\newcommand{\eea}{\end{eqnarray}}
\newcommand{\beqar}{\begin{eqnarray}}
\newcommand{\eeqar}[1]{\label{#1} \end{eqnarray}}
\newcommand{\pleft}{\stackrel{\leftarrow}{\partial}}
\newcommand{\pright}{\stackrel{\rightarrow}{\partial}}
\newcommand{\eq}[1]{Eq.~(\ref{#1})}
\newcommand{\fig}[1]{Fig.~\ref{#1}}
\newcommand{\eff}{ef\!f}
\newcommand{\alphas}{\alpha_s}

\newcommand{\jx}[1]{{\color{blue}[\textbf{JX:}\,{#1}]}}

\title{Bayesian Inference of Heavy-Quark Dissipation and Jet Transport Parameters from D-Meson observables in heavy-ion collisions at the LHC energies}

\date{\today  \hspace{1ex}}
\author{Xu-Fei Xue}
\affiliation{School of Mathematics and Physics, China University of Geosciences, Wuhan 430074, China }

\author{Zi-Xuan Xu}
\affiliation{Key Laboratory of Quark \& Lepton Physics (MOE) and Institute of Particle Physics, \\Central China Normal University, Wuhan 430079, China}

\author{Wei Dai\footnote{weidai@cug.edu.cn}}
\email{corresponding author: weidai@cug.edu.cn}
\affiliation{School of Mathematics and Physics, China University of Geosciences, Wuhan 430074, China }

\author{Jiaxing Zhao}
\affiliation{Helmholtz Research Academy Hesse for FAIR (HFHF), GSI Helmholtz Center for Heavy Ion Physics, Campus Frankfurt, 60438 Frankfurt, Germany}
\affiliation{Institut f\"ur Theoretische Physik, Johann Wolfgang Goethe-Universität, Max-von-Laue-Straße 1, D-60438 Frankfurt am Main, Germany}

\author{Ben-Wei Zhang}
\affiliation{Key Laboratory of Quark \& Lepton Physics (MOE) and Institute of Particle Physics, \\Central China Normal University, Wuhan 430079, China}

\begin{abstract}

We perform the first simultaneous Bayesian inference of the temperature-dependent heavy-quark spatial diffusion coefficient $2\pi T\mathcal{D}_s$ and the scaled jet transport coefficient $\hat{q}/T^3$ in the quark-gluon plasma, utilizing $D$-meson nuclear modification factor $R_\text{AA}$ and elliptic flow $v_2$ data from Pb-Pb collisions at $\sqrt{s_\text{NN}} = 5.02\ \text{TeV}$. The analysis employs a unified improved Langevin transport model that incorporates both collisional and radiative energy loss, followed by coalescence plus fragmentation hadronization. The posterior distributions of the parameters of $\hat{q}/T^3$ and those of $2\pi T\mathcal{D}_s$ are well constrained, and compared with the results of theoretical models or other experimental data extraction, respectively. The $30-50\%$ centrality data provide significantly stronger constraints than the $0-10\%$ data. The extracted ratio $\hat{q}/\kappa$ between the quark jet transport and heavy-quark diffusion coefficients exhibits a non-monotonic temperature dependence, deviating from the value $2$ estimated from the definition, with a value interval spanning 0.25--0.8 corresponding to the mean values of the inferred parameters. This work establishes a data-driven quantitative relationship between these two fundamental transport properties in the same observables, offering crucial insight into their interplay in the strongly coupled medium.
\end{abstract}

\pacs{???}

\maketitle

\section{Introduction}
Understanding the properties of the quark-gluon plasma (QGP) is a primary goal of high-energy heavy-ion collisions. Charm quarks are ideal probes for this purpose: produced in initial hard collisions before the QGP formation and with masses exceeding the temperature of the medium, they propagate independently. Their subsequent in-medium evolution and hadronization processes, which encode a record of parton-medium interactions, establish charm meson observables as essential hard probes of the QGP. The ongoing Run III at the LHC is delivering data with unprecedented precision, enabling highly accurate measurements of heavy-flavor production and flow~\cite{Torres:2025cic,Wu:2025byb}. These advancements provide a powerful tool for probing the initial nuclear state and for detailed studies of the QGP evolution. Furthermore, a precise description of in-medium heavy-quark transport and the development of accurate hadronization mechanisms are mutually reinforcing, driving a deeper, self-consistent understanding of heavy meson production. 

The evolution of heavy quarks in the hot, dense medium is commonly described by transport models, such as the Boltzmann
~\cite{Zhang:2005ni,Molnar:2006ci,Uphoff:2011ad,Uphoff:2012gb,Gossiaux:2010yx,Nahrgang:2013saa,Das:2013kea,Cao:2016gvr,Cao:2017hhk,Rapp:2018qla,Ke:2018jem} 
and Langevin equations
~\cite{Svetitsky:1987gq,Moore:2004tg,Mustafa:2004dr,vanHees:2004gq,vanHees:2005wb,Cao:2011et,Lang:2012nqy,Cao:2013ita,He:2013zua,Beraudo:2014boa,Das:2016llg,Scardina:2017ipo,He:2019vgs,Li:2019lex,Dai:2018mhw,Wang:2019xey,Wang:2020ukj}. Traditional implementations of these frameworks primarily incorporate collisional energy loss. To unify the treatment of both elastic and in-medium radiative energy loss, gluon radiation processes are now integrated into the dynamical evolution. In the Langevin formalism, the collisional energy loss strength is governed by the spatial diffusion coefficient $D_\text{s}$ (usually with the form $2\pi T\mathcal{D}_s$), whose temperature dependence has been extensively studied by lattice QCD~\cite{Banerjee:2011ra,Francis:2015daa,Brambilla:2020siz,Altenkort:2020fgs,Brambilla:2022xbd,Banerjee:2022gen,Altenkort:2023oms,Altenkort:2023eav,HotQCD:2025fbd}, holographic models~\cite{Policastro:2002se,Casalderrey-Solana:2006fio,Gubser:2006qh,Horowitz:2015dta,Andreev:2017bvr,Grefa:2022sav,Chen:2024epd,Dai:2025dir}, T-matrix~\cite{Mannarelli:2005pz,Riek:2010fk,Liu:2016ysz,Liu:2017qah,Tang:2023tkm}, DQPM~\cite{Cassing:2008nn,Song:2015sfa,Berrehrah:2016vzw}, QPM~\cite{Plumari:2011mk,Das:2015ana,Sambataro:2023tlv}, and also Bayesian result from Duke~\cite{Xu:2017obm}. On the other hand, the temperature-dependent jet transport coefficient $\hat{q}$, which governs radiative energy loss, has been constrained by jet quenching models using both light hadron nuclear modification factors $R_\text{AA}$ and $v_2$ and jet observables at higher transverse momentum~\cite{Chen:2010te,Chen:2011vt,JET:2013cls,Liu:2015vna,Soltz:2019aea,JETSCAPE:2021ehl,JETSCAPE:2021rfx,JETSCAPE:2024cqe,Xie:2019oxg,Xie:2024xbn}. 

The heavy quark dissipation coefficient $\kappa$ (related to $2\pi T\mathcal{D}_s$) and jet transport coefficient $\hat{q}$ share the common physical origin of transverse momentum broadening, and are approximately related as $\rm 2C_A/C_F$ ($=2$ for jet transport coefficient) estimated from the definition~\cite{JET:2013cls,Li:2021xbd,Li:2024wqq}.  A more precise delineation of their relationship would not only help to quantitatively disentangle the contributions of collisional and radiative energy loss mechanisms across different physical processes and observables, to provide a more reliable transport model as a baseline for studying hadronization, but would also help to constrain various theoretical models. For example, calculations based on the AdS/CFT correspondence provide predictions for $\hat{q}$~\cite{Liu:2006ug} and $2\pi T\mathcal{D}_s$~\cite{Gubser:2006qh} separately; in its strong-coupling limit, AdS/CFT gives $\hat q/\kappa={\sqrt{\pi}\Gamma[3/4]\over \Gamma[5/4]}\approx 2.4$. A direct extraction of their explicit relationship， taking into account their temperature dependence, using multiple observables from a single hadron species within a unified Langevin framework, has not yet been pursued. Critical questions regarding the precise interplay between the two parameters and a potential temperature dependence of their relationship are, to date, unanswered and constitute the primary motivation for this work.

Therefore, in this work, we employ the well-established hadronization model combining coalescence and fragmentation from previous work~\cite{Xu:2025ivv}. Using a state-of-the-art Bayesian inference framework with the latest published experimental Data regarding to charm meson~\cite{CMS:2020bnz,ALICE:2021rxa, ALICE:2021kfc}, we simultaneously infer the temperature dependence of the two key parameters governing in-medium transport: the dissipation coefficient $2\pi T\mathcal{D}_s$ and the jet transport coefficient $\hat{q}$, further, quantitatively determine the interplay between these fundamental quantities. Accordingly, this paper is structured as follows. In Section II, we present the theoretical framework for heavy-quark Langevin dynamics and energy loss. We also describe the parameterization of the two key parameters. Our hadronization approach is detailed in Section III. Section IV introduces the Bayesian inference methodology and presents our main results. We conclude with a summary in Section V.

\section{In-medium evolution}
 We begin with the assumption that all of the charm quarks are produced in initial hard scatterings, and the number of charm quarks is conserved in the subsequent evolutions. The initial spatial distribution of nucleon participants and binary collisions is first simulated using the Monte Carlo Glauber model~\cite{Miller:2007ri}. The production vertices of charm quarks are then sampled according to the local density of binary nucleon-nucleon collisions. The initial momentum distribution of the charm quarks is obtained from the fixed-order-next-to-leading-log (FONLL) calculation~\cite{Cacciari:2012ny}. To simulate the charm-quark propagation in hot and dense medium, the improved SHELL model~\cite{Wang:2019xey,Dai:2018mhw,Wang:2020ukj,Wang:2020qwe} is adopted, which includes collisional energy loss and medium-induced gluon bremsstrahlung. It has provided adequate descriptions of heavy-flavor jet modifications~\cite{Cao:2013ita,Wang:2019xey,Dai:2018mhw,Wang:2020ukj,Wang:2020qwe}. In the SHELL model, the heavy quarks are treated as classical particles due to their large masses, and their momentum and position evolution within a single time step are described by a modified Langevin equation:
 \begin{eqnarray}
	dx_j & = & \dfrac{p_j}{E}dt ,\label{eqn:dxj} \\ 
	dp_j & = & -\Gamma(p)p_jdt+\sqrt{\kappa(p)dt}\rho_j -\vec{p}_g. 
    \label{eqn:dpj}
\end{eqnarray}
$j$ represents three directions in the local rest frame. $\Gamma(p)$ is the drag coefficient that characterizes the strength of the viscous drag force. $\sqrt{\kappa(p)dt}\rho_j$ describes the stochastic momentum fluctuations of charm quarks caused by random kicks from thermal quarks and gluons in the QGP medium. $\rho_j$ follows a normal distribution $P(\vec{\rho}) = (2\pi)^{-3/2}e^{-\vec{\rho}^2/2}$.
$\kappa(p)$ is the momentum diffusion coefficient, which is related to the spatial diffusion coefficient $\mathcal{D}_s$ and the drag coefficient through the classic fluctuation-dissipation theorem~\cite{Kubo:1966fyg}:
\begin{eqnarray}
	\kappa=2\Gamma ET=\dfrac{2T^2}{\mathcal{D}_s}.
    \label{eqn:fluc-diss}
\end{eqnarray}
To account for the temperature dependence of the spatial diffusion coefficient $\mathcal{D}_s$, we adopt a linear relationship between $2\pi T \mathcal{D}_s$ and the normalized temperature $T/T_c$ (with $T_c$ denoting the critical temperature), following the lattice QCD results of Ref. \cite{Altenkort:2023oms}. This relationship is expressed as the slope-intercept form of a linear equation:

\begin{equation}
2\pi T \mathcal{D}_s = k \cdot \frac{T}{T_c} + b,
\label{eq:ds-t}
\end{equation}
where slope $k$ and y-intercept $b$ are free model parameters to be constrained by the analysis presented in this work. To benefit further Bayesian analysis, they are denoted as $k_{\rm 2\pi T \mathcal{D}_s}$ and $b_{\rm 2\pi T \mathcal{D}_s}$. 

The term $\vec{p}_g$ in Eq.~(\ref{eqn:dpj}) manifests the energy loss attribute to medium-induced gluon radiation. The radiated gluon momentum spectra are derived from the higher-twist energy loss formalism~\cite{Guo:2000nz,Zhang:2003wk,Zhang:2004qm,Majumder:2009ge}, the form reads
\begin{eqnarray}
    \dfrac{dN}{dxdk^2_{\perp}dt}=\dfrac{2\alpha_sC_sP(x)\hat{q}}{\pi k^4_{\perp}}\sin^2(\dfrac{t-t_i}{2\tau_f})(\dfrac{k^2_{\perp}}{k^2_{\perp}+x^2M^2})^4. \label{eqn:gluon_radiation_spectra}
\end{eqnarray}
$\alpha_s$ is the strong coupling constant. $C_s$ is the quadratic Casimir in color representation. $P(x)$ is the vacuum splitting function\cite{Deng:2009ncl}. $k_{\perp}$ and $x$ represent the transverse momentum and the energy fraction of the parent parton carried by the emitted gluon, respectively. $(t-t_i)$ represents the time interval between two successive gluon emissions. $\tau_f=2Ex(1-x)/(k^2_{\perp}+x^2M^2)$ denotes the formation time of the emitted gluon. $\rm M$ is the mass of the parent quark.
$\hat{q}$ is the jet transport parameter. We assume the jet transport coefficient $\hat{q}$ scales with the cube of the local quark-gluon plasma (QGP) temperature $T$ ($\hat{q} \propto T^3$). We adopt a linear interpolation ansatz for $\hat{q}/T^3$ between the initial temperature $T_0$ and the QCD critical temperature $T_\mathrm{c}$. The dependence is expressed as the same two-point form of the linear parametrization~\cite{Xie:2024xbn}:

\begin{equation}
\frac{\hat{q}}{T^3} = \left[ \left( \frac{\hat{q}_0}{T_0^3} - \frac{\hat{q}_\mathrm{c}}{T_\mathrm{c}^3} \right) \frac{T - T_\mathrm{c}}{T_0 - T_\mathrm{c}} + \frac{\hat{q}_\mathrm{c}}{T_\mathrm{c}^3} \right] \frac{p^\mu \cdot u_\mu}{p_0},
\label{eq:qhat-T}
\end{equation}
where $\hat{q}_0$ denotes the initial value of the jet transport coefficient at the initial temperature $T_0$ at the most central position of the fireball (provided by the hydrodynamic model), whose value depends on the collision system and can be found in Table \ref{tab:hydro-initial-T}, however for the Data combined analysis, it means the largest initial temperature within all system involved. $\hat{q}_\mathrm{c}$ denotes the value of the jet transport coefficient at the critical temperature $T_\mathrm{c}$, taken as 0.165 GeV/c in our study. The local temperature $T$ is constrained to the range $T \in [T_\mathrm{c}, T_0]$ and $p^\mu$ is the parton four-momentum, $u_\mu$ is the medium four-velocity, and $p_0$ is a reference momentum for normalization.
\begin{table}[htbp]
  \centering
  \caption{Initial temperatures $ T_0$ for different centrality classes in Pb-Pb collisions at $ \sqrt{s_{\rm NN}} = 5.02 $ TeV.}
  \label{tab:hydro-initial-T}
  \begin{tabular}{cc}
    \toprule
    Centrality Class & $ T_0$ (GeV) \\
    \midrule
    0--10\%       & 0.505 \\
    30--50\%  & 0.445 \\
    \bottomrule
  \end{tabular}
\end{table}

By integrating the gluon radiation spectra, one gets the average number of emitted gluons at a time step. Then, assuming that the gluon radiation process obeys a Poisson distribution, the actual number of radiated gluons can be determined. The momentum of the emitted gluon is sampled through Eq.~(\ref{eqn:gluon_radiation_spectra}). 

Because the QGP medium exhibits fluid-like behavior, its spacetime evolution is described by the equations of viscous hydrodynamics. We utilize the (3+1)-dimensional CLVisc hydrodynamics model~\cite{Pang:2018zzo,Wu:2021fjf} to simulate the dynamical evolution of QGP and extract the fluid velocity and local temperature at each time step. 

\section{Hadronization}
When a charm quark has propagated through the medium and the local temperature of its fluid cell reaches a critical value, it is assumed to undergo hadronization simultaneously. We incorporate a coalescence mechanism for lower momentum and a fragmentation function for higher momentum to simulate the hadronization of charm quarks. The coalescence mechanism, in which hadrons are formed by the combination of their constituent quarks in close proximity in phase space, was proposed to explain the unexpectedly high yield ratio enhancement and scaled elliptic flow observed in experiment~\cite{Hwa:2002tu,Greco:2003xt,Fries:2003kq,Molnar:2003ff}. The momentum distribution of hadrons produced from quark coalescence at the hadronization hypersurface is given by
\begin{eqnarray}
	\dfrac{dN_h}{d^2p_Td\eta} = &c \displaystyle{\int} p^{\mu}d\sigma_{\mu} \prod \limits_{i=1}^n \dfrac{d^4x_id^4p_i}{(2\pi)^3}f_i(x_i,p_i) \\ \notag
                   &\times g_H\cdot W_h(x_1,\cdots , x_i, p_1,\cdots , p_i).
                   \label{eqn:coal_distribution}
\end{eqnarray}
$i$ represents the index of each constituent quark. $g_H$ is the spin-color factor. The parameter $c$ is a normalization factor, which ensures that the sum of the momentum spectra of all considered charmed hadron states equals unity when the momentum of the charm quark approaches zero. This condition reflects the assumption that under such circumstances, all charm quarks undergo hadronization exclusively via coalescence. $f_i(x_i,p_i)$ is the phase-space distribution function of each constituent quark at the hadronization hypersurface. For thermal-light quarks, the momentum distribution is $f_q(r_i,p_i)=N_q/(e^{u_{\mu}p^{\mu}_i/T}+1)$ with the degeneracy factor $N_q = 6$. 
$W_h$ is the Wigner density, which is conventionally considered as the probability amplitude for the quark coalescence; it can be obtained via the Wigner transformation of the hadronic wave function of a harmonic oscillator potential. Take charmed mesons as an example, the phase-space Wigner function of each wave state involved in this model reads~\cite{Zhao:2025cnp}
\begin{eqnarray}
W_{1S}({\bf r},{\bf p})&=&8e^{-\xi},\nonumber\\
W_{1P}({\bf r},{\bf p})&=&\dfrac{8}{3}e^{-\xi}(2\xi - 3),\nonumber\\
W_{1D}({\bf r},{\bf p})&=&\dfrac{8}{15}e^{-\xi}(15+4\xi^2-20\xi+8\eta),\nonumber\\
W_{2S}({\bf r},{\bf p})&=&\dfrac{8}{3}e^{-\xi}(3+2\xi^2-4\xi-8\eta),\nonumber\\         
W_{2P}({\bf r},{\bf p})&=&\dfrac{8}{15}e^{-\xi}(-15+4\xi^3-22\xi^2+30\xi \nonumber\\ 
&-&8(2\xi-7)\eta,
\label{eqn:normal1}
\end{eqnarray}
where $\xi \equiv r^2/\sigma^2+p^2\sigma^2$ and $\eta = p^2r^2-({\bf p}\cdot {\bf r})^2$. $r$ and $p$ are the relative coordinate and relative momentum of two constituent quarks in their center-of-mass (CM) frame, respectively. For charmed mesons, the relative momentum is given by
\begin{eqnarray}
p&=&\dfrac{m_1\cdot p_c - m_c\cdot p_1}{m_1+m_c}.
\label{eqn:rel_pm}
\end{eqnarray}
For charmed baryons, there are two relative momenta:
\begin{eqnarray}
p_{\rho}&=&\dfrac{m_1\cdot p_c - m_c\cdot p_1}{m_1+m_c},\nonumber\\
p_{\lambda}&=&\dfrac{m_2\cdot (p_c+p_{1})-(m_c+m_1)\cdot p_{2}}{m_1+m_2+m_c},
\label{eqn:rel_pb}
\end{eqnarray}
where $p_c$ and $m_c$ are charm-quark momentum and mass in the center-of-mass frame, $p_i$ and $m_i$ are the i-th light-quark momentum and mass in the CM frame. For charmed baryons, we assume a two-step process: the charm quark first combines with a light constituent quark to form a charm-light system, which subsequently coalesces with the second light quark as an entirety. Thus, the baryon has two center-of-mass frames. 
For charmed mesons, the Gaussian width $\sigma$ can be determined by the average radius as:
\begin{eqnarray}
    \langle r \rangle = \int \dfrac{d^3rd^3p}{(2\pi)^3}W(r,p)r
    \label{eqn:width_meson}
\end{eqnarray}
and for charmed baryons,
\begin{eqnarray}
    \langle \rho \rangle = \dfrac{d^3\rho d^3\lambda d^3p_{\rho}d^3p_{\lambda}}{(2\pi)^6}W(\rho,\lambda,p_{\rho},p_{\lambda}) \rho,\nonumber \\
    \langle \lambda \rangle = \dfrac{d^3\rho d^3\lambda d^3p_{\rho}d^3p_{\lambda}}{(2\pi)^6}W(\rho,\lambda,p_{\rho},p_{\lambda}) \lambda.
    \label{eqn:width_baryon}
\end{eqnarray}
The averaged radius of the ground state of charmed mesons and baryons has been calculated by solving the Dirac equation with a lattice quark potential, as described in Refs.~\cite{Shi:2013rga,Shi:2019tji}. 
To simplify the calculation, we integrate out the coordinate space degrees of freedom and retain only the momentum-space distribution.

We calculate the total coalescence probability as a function of charm-quark momentum by Eq.~(\ref{eqn:coal_distribution}) to decide if a charm quark hadronizes through the coalescence process when it reaches the hadronization hypersurface. 
Once a random number satisfies this probabilistic condition, light quarks are sampled from the thermal medium, and the relative coalescence probability for each state is calculated to determine the corresponding hadronic final state. 
If the charm quark cannot coalesce, we let it hadronize through fragmentation, the momentum of the final-state hadron is given by the Peterson fragmentation function~\cite{Peterson:1982ak}
\begin{eqnarray}
	\mathscr{D}_{c\to H}(z) \propto \dfrac{1}{z\left(1-\dfrac{1}{z}-\dfrac{\epsilon}{1-z}\right)^2},
	\label{eqn:Peterson}
\end{eqnarray}
where $z$ is the momentum fraction of a charmed hadron fragmented from a charm quark. We take parameter $\epsilon = 0.01$ for charmed mesons and $\epsilon = 0.02$ for charmed baryons~\cite{Das:2016llg}.
We refer to the Charm-quark fragmentation fractions in hadrons of experimental data from $\rm p+p$ collisions~\cite{Lisovyi:2015uqa} to determine the charmed-hadron final states produced by fragmentation.

We also considered the D-meson hadronic rescattering process by using the Langevin equation. In the hadronic phase, the diffusion coefficient exhibits a strong dependence on temperature. We employ the temperature-dependent diffusion coefficient from Ref.~\cite{Torres-Rincon:2021yga} for hadronic interaction of D-meson, with the kinetic freeze-out temperature fixed at 137~MeV~\cite{Pang:2018zzo}.

\section{Bayesian Inference}\label{sec3}

\begin{figure}[htbp]
\centering
\includegraphics[width=0.50\textwidth]{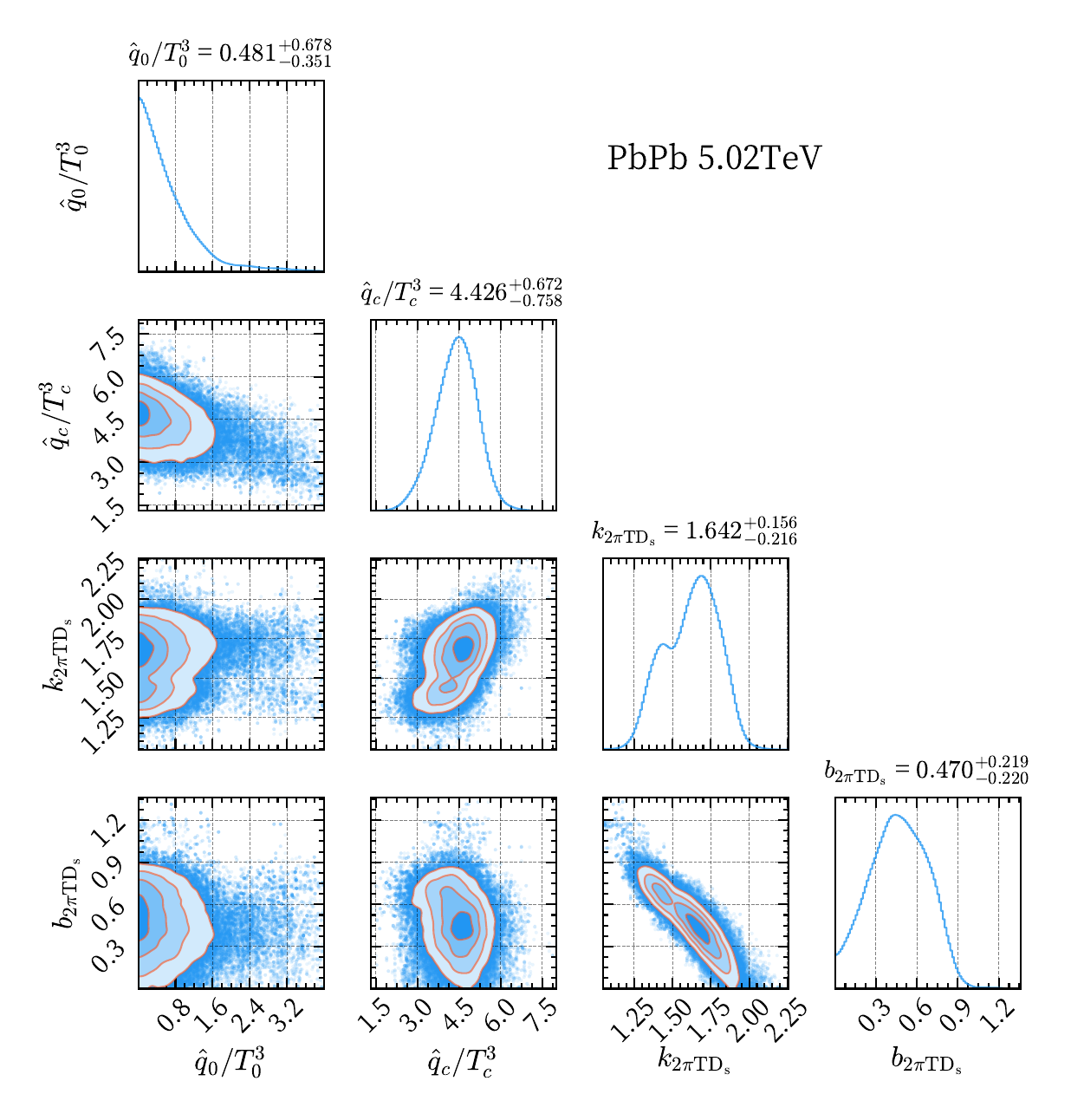}
\caption{Posterior distributions (diagonal panels) and pairwise correlations (off-diagonal panels) of the model parameters $\hat{q}_0/T_0^3$, $\hat{q}_\mathrm{c}/T_\mathrm{c}^3$, $k_{\rm 2\pi T \mathcal{D}_s}$, and $b_{\rm 2\pi T \mathcal{D}_s}$ for Pb+Pb collisions at $\sqrt{s_{\rm NN}} = 5.02$ TeV. All panels display results with 95\% credibility intervals. Combined analysis of 0--10\% and 30--50\% centrality classes.}
\label{fig:param-posterior-corner}
\end{figure}

We employ a hierarchical Bayesian inference framework (implemented using PyMC) to investigate the probabilistic distributions and correlations of crucial QGP transport parameters, including the jet transport coefficient $\hat{q}$ and the spatial diffusion coefficient $\mathcal{D}_s$.

Based on Eqs. \ref{eq:qhat-T} and \ref{eq:ds-t}, our model involves four key parameters to be inferred, as listed in Table \ref{tab:prior-params}. 

\begin{table}[htbp]
  \centering
  \caption{Prior ranges of the model parameters}
  \label{tab:prior-params}
  \begin{tabular}{cc}
    \toprule
    Parameter & Prior Range \\
    \midrule
    $\rm \hat{q}_0/T_0^3$ & 0--6.0 \\
    $\rm \hat{q}_\mathrm{c}/T_\mathrm{c}^3$ & 0--9.0 \\
    $k_{\rm 2\pi T \mathcal{D}_s}$ & 0.5--3.0 \\
    $b_{\rm 2\pi T \mathcal{D}_s}$ & 0--4.5 \\
    \bottomrule
  \end{tabular}
\end{table}

All priors are assumed to follow uniform distributions to avoid overly strong prior assumptions. Their ranges are determined based on physical constraints from lattice QCD calculations \cite{Altenkort:2023oms} and previous studies \cite{Xie:2024xbn}.

The likelihood is assumed to take a skewed normal form:

\begin{equation}
\mathcal{L}(\mathrm{data} \mid \theta) = \prod_i \mathcal{SN}\left(\mathrm{data}_i \mid \mu_i(\theta), \sigma_i\right),
\label{likelihood}
\end{equation}
where data represents experimental data and $\theta$ represents the key model parameters listed in Table \ref{tab:prior-params}; $\mathcal{SN}$ denotes the skewed normal distribution, introduced to account for the asymmetric uncertainties of the experimental data;  $\mu_i$ denotes the theoretical predictions generated by a surrogate model; and $\sigma_i$ encompasses both the experimental uncertainties of individual data points and the uncertainties associated with the theoretical predictions. The surrogate model is constructed via regular-grid interpolation, using precomputed theoretical simulations from the SHELL model that span the entire parameter space.

Through Bayes' theorem, the posterior distribution is proportional to the product of the prior distributions and the likelihood function:

\begin{equation}
P\left(\theta \mid \mathrm{data}\right) \propto \left( \prod_i P_i\left(\theta_i\right) \right) \mathcal{L}\left(\mathrm{data} \mid \theta\right),
\label{posterior}
\end{equation}
where $\theta$ collectively denotes the model parameters, $P_i(\theta_i)$ are the individual prior distributions for each parameter $\theta_i$, and $\mathcal{L}(\mathrm{data} \mid \theta)$ is the likelihood function. Posterior distributions are sampled using the Metropolis-Hastings algorithm within a Markov Chain Monte Carlo (MCMC) framework \cite{andrieu2003introduction}. Convergence is assessed using the $\hat{R}$ statistic, with all parameters yielding values below 1.01, thereby ensuring reliable convergence. Additionally, posterior predictive checks are performed to validate the compatibility of the theoretical model with experimental data.
Using the hierarchical Bayesian inference framework described above, we perform a parameter extraction analysis for the parameters listed in Table \ref{tab:prior-params} with broad ranges of prior values, based on multiple experimental observables within D-mesons in Pb-Pb collisions at $\sqrt{s_{\rm NN}} = 5.02$ TeV. These observables include the transverse momentum spectra ($d^2N/dp_Tdy$), nuclear modification factors ($R_\text{AA}$), elliptic flow coefficients ($v_2$), and the particle ratio ($D_s^+/D^0$) for $D^0$ and $D_s^+$ mesons. Details of the observables, their corresponding centrality classes, and the experimental references are summarized in Table.~\ref{tab:observables}.
\begin{table}[htbp]
  \centering
  \caption{Experimental observables used in the likelihood function.}
  \label{tab:observables}
  \begin{tabular}{cccc}
    \toprule
    Centrality Class & Observable                          & Particle       & Reference       \\
    \midrule
    \multirow{4}{*}{0--10\%} & $d^2N/dp_Tdy$ & $D^0$, $D_s^+$         & ALICE\cite{ALICE:2021rxa, ALICE:2021kfc} \\
                             & $R_\text{AA}$                           & $D^0$                  & ALICE\cite{ALICE:2021rxa}  \\
                             & $v_2$                              & $D^0$                  & CMS\cite{CMS:2020bnz}      \\
                             & $D_s^+/D^0$                   & $D^0$, $D_s^+$    & ALICE\cite{ALICE:2021kfc}  \\
    \midrule
    \multirow{4}{*}{30--50\%} & $d^2N/dp_Tdy$& $D^0$, $D_s^+$         & ALICE\cite{ALICE:2021rxa, ALICE:2021kfc} \\
                              & $R_\text{AA}$                           & $D^0$                  & ALICE\cite{ALICE:2021rxa}  \\
                              & $v_2$                              & $D^0$, $D_s^+$         & CMS\cite{CMS:2020bnz}, ALICE\cite{ALICE:2021kfc} \\
                              & $D_s^+/D^0$                  & $D^0$, $D_s^+$    & ALICE\cite{ALICE:2021kfc}  \\
    \bottomrule
  \end{tabular}
\end{table}

Fig.~\ref{fig:param-posterior-corner} presents the posterior distributions (diagonal panels) and pairwise correlations (off-diagonal panels) for the model parameters $\hat{q}0/T_0^3$, $\hat{q}\mathrm{c}/T_\mathrm{c}^3$, $k_{\rm 2\pi T \mathcal{D}_s}$, and $b_{\rm 2\pi T \mathcal{D}_s}$ extracted from the experimental Data for both $0-10\%$ and $30-50\%$ centrality Pb+Pb collisions at $\sqrt{s_{\rm NN}} = 5.02$ TeV. Mean values of all the parameters, including $95\%$ credibility intervals, are listed above the corresponding diagonal panels. $\hat{q}_0/T_0^3$, $\hat{q}_\mathrm{c}/T_\mathrm{c}^3$, $k_{\rm 2\pi T \mathcal{D}_s}$, and $b_{\rm 2\pi T \mathcal{D}_s}$ are constrained with distinct peaks in their posterior distributions. Based on the current inferred credibility intervals, the estimation for $\hat{q}_0$ exhibits the poorest precision, with its posterior distribution skewed toward zero. The parameter $b_{\rm 2\pi T \mathcal{D}_s}$ has the next-lowest precision. In contrast, the jet transport coefficient at $T_c$ and the slope of the spatial diffusion coefficient are better constrained, both with credibility intervals within ±15\% of their mean values. We performed Bayesian inference for these four parameters separately within two centrality classes. The credibility intervals obtained from the 0-10\% centrality data alone are notably wider than those derived from the 30-50\% data. The posterior means from the combined analysis of all data are closer to those from the 30-50\% centrality class, indicating the experimental Data sets from the 30-50\% centrality class constrain models better than those from 0-10\%. The precision of the overall inference, as reflected by the credibility intervals, is improved compared to the separate analyses. 

\begin{figure}[htbp]
\centering
\includegraphics[width=0.48\textwidth]{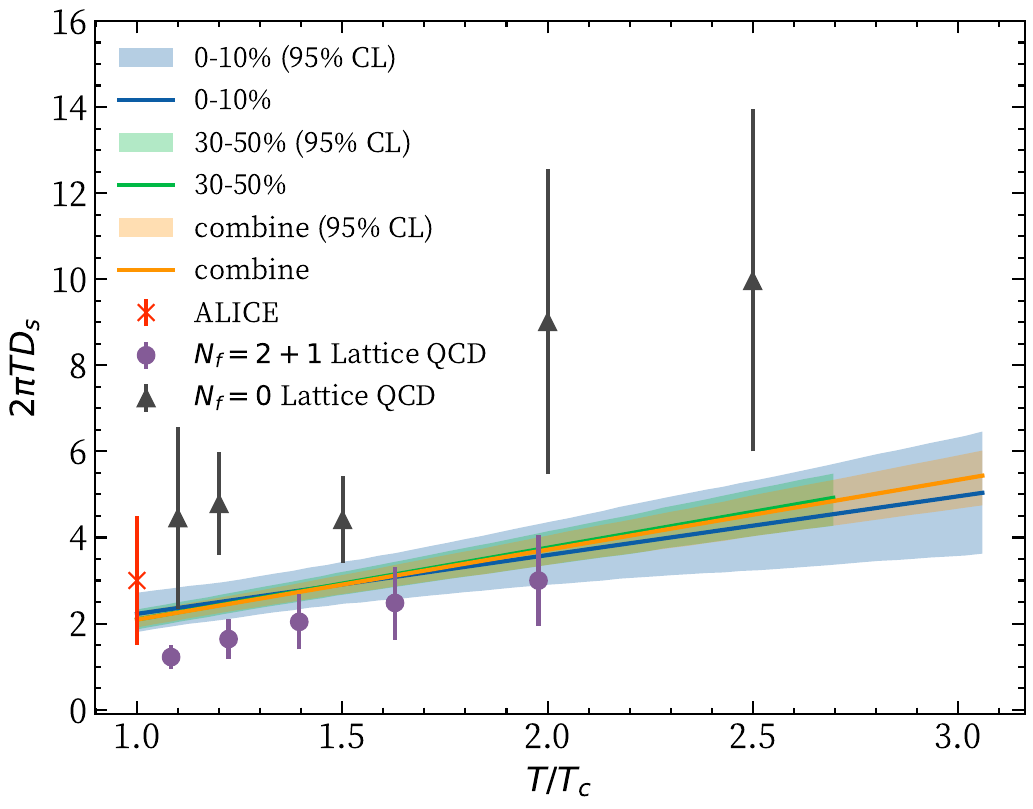}
\caption{Bayesian-inferred temperature dependence of $2\pi T\mathcal{D}_s$ as a function of $T/T_c$, comparing results obtained from the 0-10\% Pb-Pb collision system (blue), the 30-50\% centrality class (green), and their combined analysis (orange). The shaded bands indicate the 95\% credibility regions, respectively. The model-to-Data fitting result of the ALICE Collaboration is presented at $T=T_c$ with large theoretical uncertainty~\cite{ALICE:2021rxa}, the 0 flavor, and the 2+1 flavor lattice QCD results~\cite{Altenkort:2023oms} are also plotted.}
\label{fig:2piTDs-T-over-Tc}
\end{figure}

The Bayesian-inferred dependence of $2\pi T\mathcal{D}_s$ on $T/T_c$ is shown in Fig.~\ref{fig:2piTDs-T-over-Tc}, presenting results separately extracted from experimental data in the 0-10\% Pb-Pb collision system, the 30-50\% centrality class, and from a combined analysis of both datasets. For each case, the functional relationship corresponding to the mean values of $k_{\rm 2\pi T \mathcal{D}_s}$ and $b_{\rm 2\pi T \mathcal{D}_s}$ is displayed, along with the shaded areas using the 95\% credibility intervals of the parameters. Furthermore, we include in the plot the Bayesian inference result from ALICE (2018) obtained with their contemporary model and data, as well as a result from lattice QCD calculations. Our combined result at $T_c$ lies slightly below the ALICE inference but systematically higher than the lattice value. The slope of the $2\pi T\mathcal{D}_s$ versus $T/T_c$ function, extracted from the combined analysis of the 0-10$\%$ and 30-50$\%$ centrality data, shows better agreement with the lattice calculation. Comparison of the slopes reveals that the slope of the $2\pi T\mathcal{D}_s$ vs. $T/T_c$ function inferred solely from the 30-50$\%$ centrality data is notably close to that obtained from the combined analysis. In addition, the narrower 95$\%$ credibility regions indicate that the 30-50$\%$ centrality data impose stronger constraints on the model parameters than the 0-10$\%$ data. Furthermore, the 95$\%$  credibility region associated with each curve is narrowest near $T_c$ and widest near the initial temperature $T_0$, indicating that the experimental data provide stronger constraints on the spatial diffusion coefficient at lower temperatures than at higher temperatures.  By comparing the $T/T_c$ ranges covered by the three curves and their associated uncertainty bands, it can be inferred that the 30-50$\%$ centrality system, which has a lower initial temperature $T_0$ (see Table~\ref{tab:hydro-initial-T}) compared to the 0--10$\%$ case, provides experimental constraints only for the functional behavior below $T_0 = 0.445$~GeV.

\begin{figure}[htbp]
\centering
\includegraphics[width=0.48\textwidth]{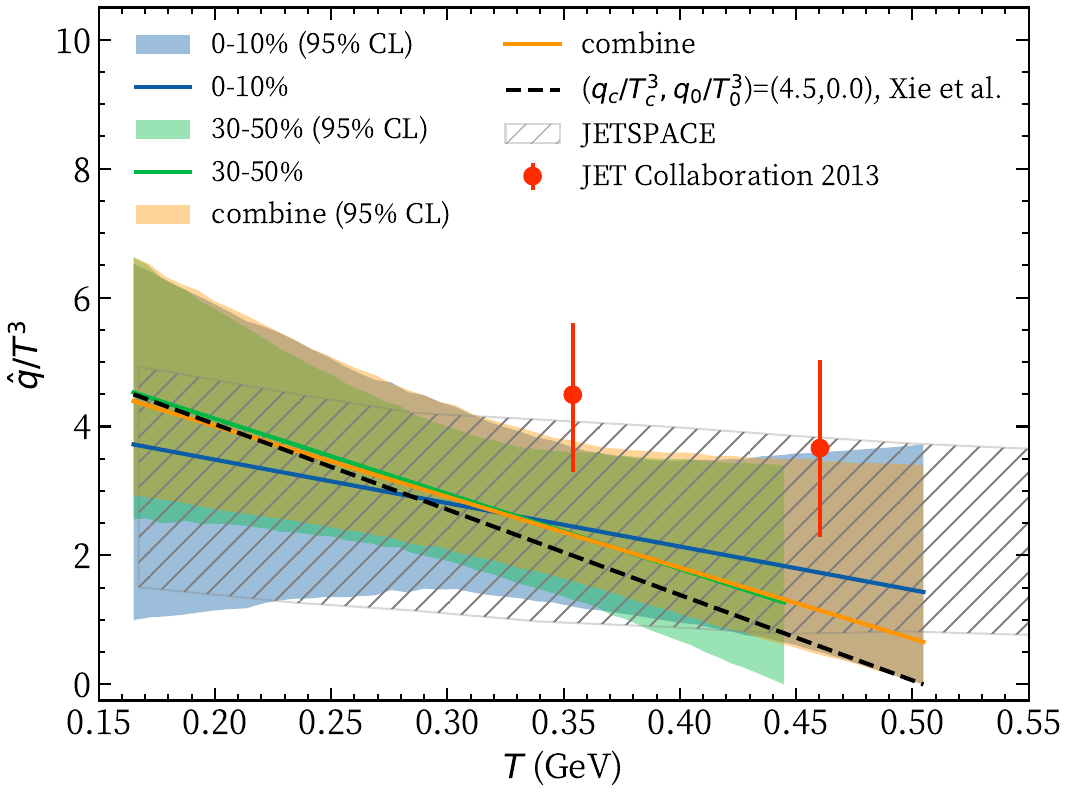}
\caption{The scaled jet transport coefficient $\hat{q}/T^3$ as a function of the medium temperature $T$ obtained from the 0-10$\%$ Pb+Pb collision system (blue), the 30-50$\%$ centrality class (green), and their combined analysis (orange), compared to results from a global fitting to experimental data of large $p_T$ light hadron at LHC (black-dashed line)~\cite{Xie:2024xbn}, 
JETSCAPE (black shadow)~\cite{JETSCAPE:2021ehl} and the JET Collaboration (red dots)~\cite{JET:2013cls}. All results are presented using mean values of the inferred parameters $\hat{q}_0/T^3$ and $\hat{q}_c/T^3$ along with 95\% credibility intervals.}
\label{fig:q-over-T3}
\end{figure}

We further present the inferred scaled jet transport coefficient $\hat{q}/T^3$ as a function of the medium temperature $T$ obeying Eq.~\ref{eq:qhat-T}, and compared it with the results from a global fit to data in A+A collisions at LHC～\cite{Xie:2024xbn}, an global extraction for JETSCAPE~\cite{JETSCAPE:2021ehl} and also those from the JET Collaboration～\cite{JET:2013cls}, as shown in Fig.～\ref{fig:q-over-T3}. The curves are plotted using the mean values of the inferred parameters, and the shaded areas are plotted using the 95$\%$ credibility intervals of the parameters. From the differences in the slopes of the three curves, it can be observed that, similar to the case of $2\pi T\mathcal{D}_s$, the 30–50$\%$ data provide stronger constraints, and the result from the combined analysis of both centrality classes nearly coincides with that from the 30–50$\%$ data alone, even though the latter only constrains the functional form up to a lower maximum temperature. Moreover, the $95\%$ credibility regions for all three cases show that constraints are tighter in the intermediate temperature range compared to both the low- and high-temperature extremes. In the low-temperature region, the 0–10$\%$ data provide weaker constraints than the 30–50$\%$ data, leading to a smaller magnitude of the inferred negative slope in the 0–10$\%$ case than in the 30–50$\%$ case.

The scaled jet transport coefficient inferred in this work is lower than the global extraction by the JET Collaboration in 2013 based on high-transverse-momentum hadron production. It is comparable in magnitude to the results from a global extraction for JETSCAPE and also from the information-field-based (IF-based) Bayesian analysis～\cite{Xie:2022ght}. The final results are consistent with those extracted from the light-hadron $R_{\text{AA}}$ and $v_2$ data shown as the black-dashed line~\cite{Xie:2024xbn}, as the same two-point form of the linear parametrization in their work is adopted. This allows a direct comparison of the extracted values at the two reference temperatures: at $T_c$, the mean value of $\hat{q}_c/T_c^3$ obtained here is 0.47, whereas their result is 0.45. The discrepancy likely originates near $T_0$, where our extraction based on $D$-meson $R_{\text{AA}}$ and $v_2$ yields a mean $\hat{q}_0/T_0^3$ of 0.48, compared to their significantly lower value 0. Furthermore, as shown in the first diagonal panel of Fig.~\ref{fig:param-posterior-corner}, the posterior distribution of $\hat{q}_0/T_0^3$ in our study also lacks a well-defined peak. This suggests that the challenge is persistent in describing heavy-flavor observables, and indicates room for improving the linear parametrization of the temperature dependence of the jet transport coefficient. In this context, the information field approach might suggest a broadly varying, non-linear temperature dependence for $\hat{q}$. 

\begin{figure}[htbp]
\centering
\includegraphics[width=0.48\textwidth]{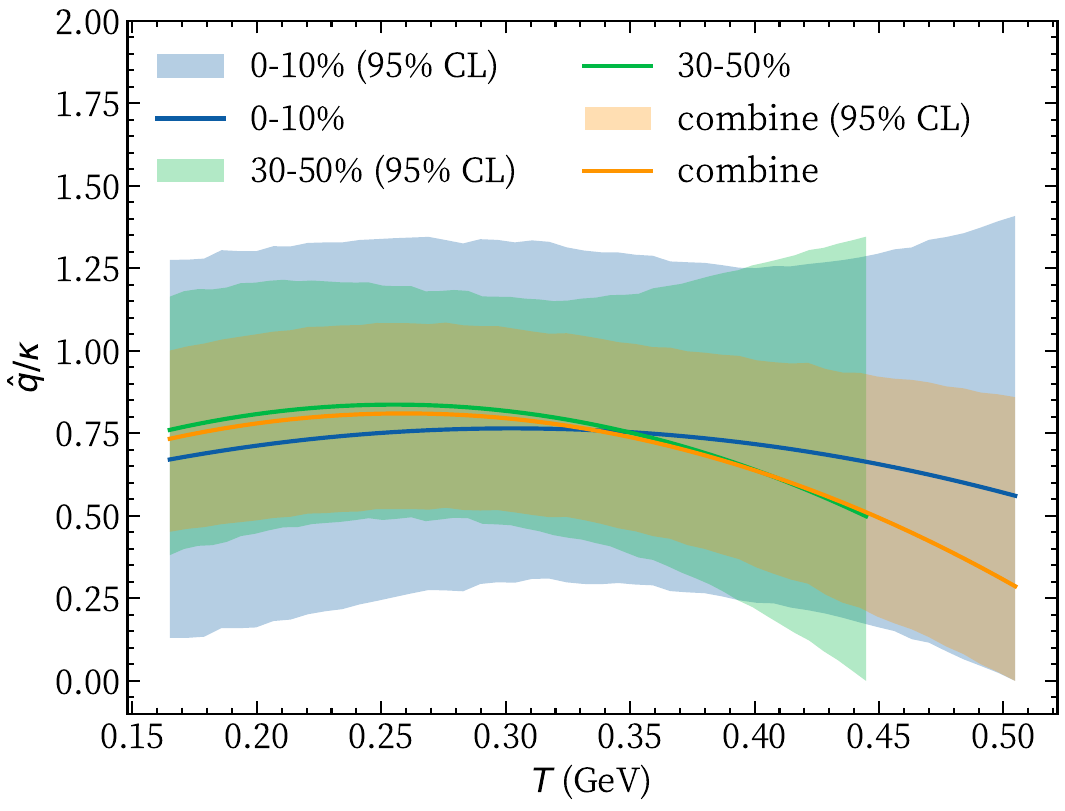}
\caption{The ratio $\hat{q}/\kappa$ as a function of the medium temperature $T$ obtained from the 0-10$\%$ Pb+Pb collision system (blue), the 30-50$\%$ centrality class (green), and their combined analysis (orange). The results are presented using mean values of the inferred parameters along with 95$\%$  credibility regions.}
\label{fig:qhat-over-kappa}
\end{figure}

Finally, after we simultaneously infer both \(\hat{q}\) and  \(2\pi T\mathcal{D}_s\) from observables in D mesons, in Fig.~\ref{fig:qhat-over-kappa} we present the inferred ratio $\hat{q}/\kappa$ of the jet transport parameter $\hat{q}$ to the heavy-quark diffusion coefficient $\kappa$, as well as its temperature dependence, obtained from the 0–10$\%$ and 30–50$\%$ centrality bins and from the combined analysis using all available D-meson $R_\text{AA}$ and $v_2$ data at LHC energy of 5.02 TeV. We first observe that the inferred ratio differs from the definition estimate of $2$, also from the strong-coupling limit of AdS/CFT results, which is around 2.4 estimated from～\cite{Gubser:2006qh,Liu:2006ug}. The 95$\%$ credibility interval spans approximately 0.1–1.25 across the studied temperature range, while the curve corresponding to the posterior mean varies between about 0.25 and 0.8 for the combined case. Its temperature dependence exhibits a gentle bump between $T_c$ and 0.35 GeV, followed by a decreasing trend when the temperature increases. 

\section{Summary}\label{sec4}

This study presents a state-of-the-art Bayesian inference of heavy-quark transport within the quark-gluon plasma (QGP) produced in Pb-Pb collisions at the LHC. By employing a unified Langevin framework that coherently incorporates both collisional and radiative energy loss mechanisms, the work simultaneously extracts the temperature dependence of two fundamental parameters: the spatial diffusion coefficient \(2\pi T\mathcal{D}_s\) and the jet transport coefficient \(\hat{q}\). The analysis utilizes the latest high-precision experimental data on \(D^0\) and \(D_s^+\) mesons---including transverse momentum spectra, nuclear modification factors \(R_{\mathrm{AA}}\), elliptic flow \(v_2\), and particle yield ratios---across 0--10\% and 30--50\% centrality classes.

One can find that data from the 30--50\% centrality class impose stronger constraints on the inferred parameters than those from the more central 0--10\% collisions. The extracted \(2\pi T\mathcal{D}_s\) near the critical temperature \(T_c\) is constrained better than at the high temperature, while its slope with temperature is consistent with first-principles predictions of LQCD. The scaled jet transport coefficient \(\hat{q}/T^3\) exhibits a temperature trend compatible with global fits based on light-flavor observables, though its value near the initial temperature \(T_0\) remains less constrained. Notably, the ratio \(\hat{q}/\kappa\) (where \(\kappa\) is the heavy-quark momentum diffusion coefficient) is found to be significantly smaller than the definition estimate of $2$, displays a non-trivial temperature dependence, decreasing from approximately 0.8 near \(T_c\) to about 0.25 at higher temperatures.

These results provide the first simultaneous, data-driven determination of the interplay between the strength parameters that govern the collisional and radiative energy losses for heavy quarks in the QGP. The work establishes a refined transport baseline for future studies of hadronization. It offers quantitative guidance for discriminating between strong-coupling theoretical models, thereby advancing our understanding of parton-medium interactions under extreme conditions.

\newpage

{\bf Acknowledgments:}  We thank Xinye Peng for helpful discussions. WD is supported by the National Key Research and Development Program of China (Grant No. 2024YFA1610804), BWZ is supported by a grant from the National Natural Science Foundation of China (Key Program) (No. 12535010), and JX is supported by the Helmholtz Research Academy Hesse for FAIR.

\section{Appendix}
\begin{figure*}[htbp]  
\centering
\begin{subfigure}[b]{0.48\textwidth}  
\centering
\includegraphics[width=\textwidth]{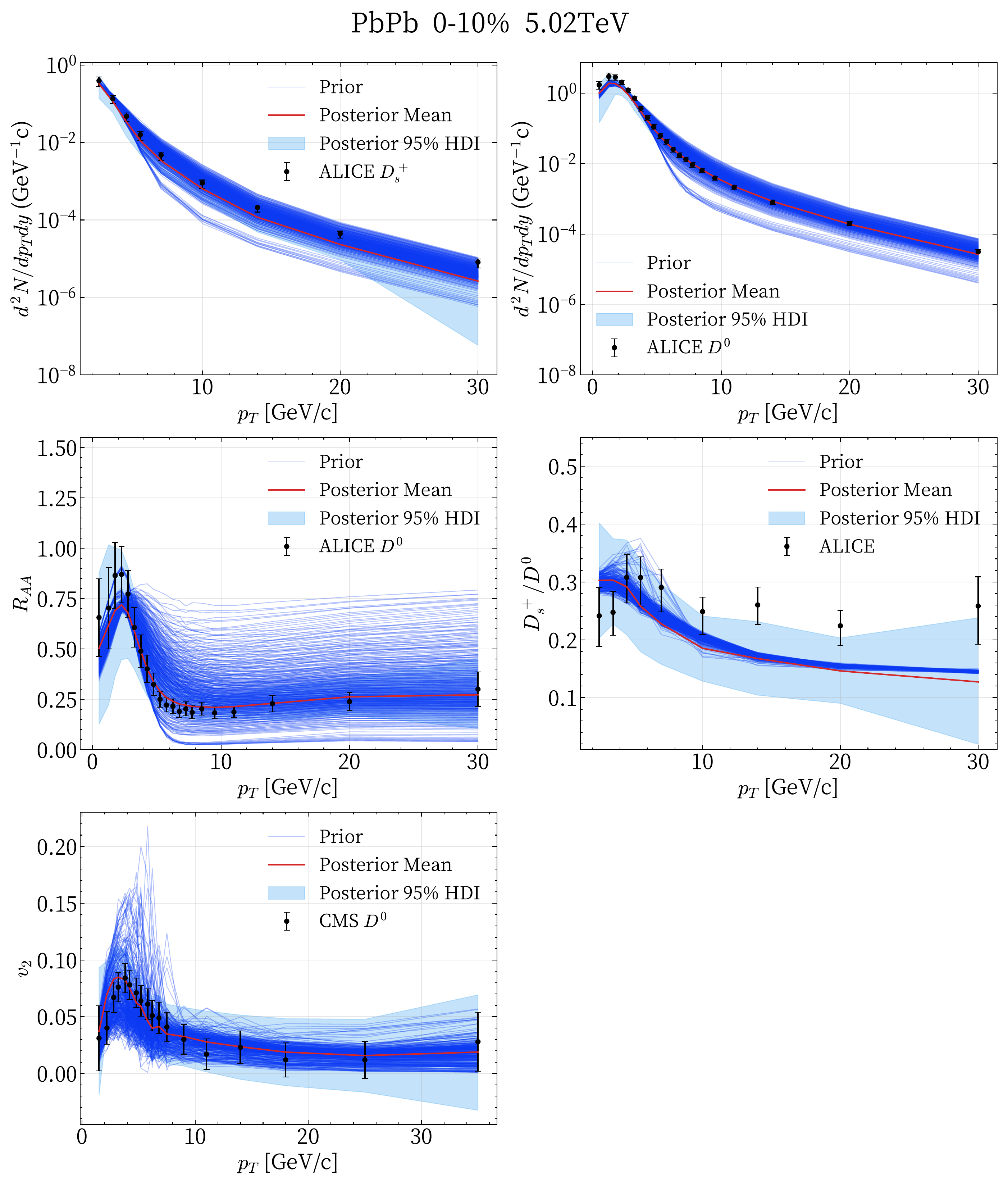}
\caption{0-10\% Pb-Pb collisions, $\sqrt{s_{NN}}=5.02$ TeV. Top-left: $D_s^+$ spectrum~\cite{ALICE:2021kfc}; Top-right: $D^0$ spectrum~\cite{ALICE:2021rxa}; Middle-left: $D^0$ $R_{AA}$~\cite{ALICE:2021rxa}; Middle-right: $D_s^+/D^0$ ratio~\cite{ALICE:2021kfc}; Bottom-left: $D^0$ $v_2$~\cite{CMS:2020bnz}}
\label{pbpb010-data}
\end{subfigure}
\hfill  
\begin{subfigure}[b]{0.48\textwidth}
\centering
\includegraphics[width=\textwidth]{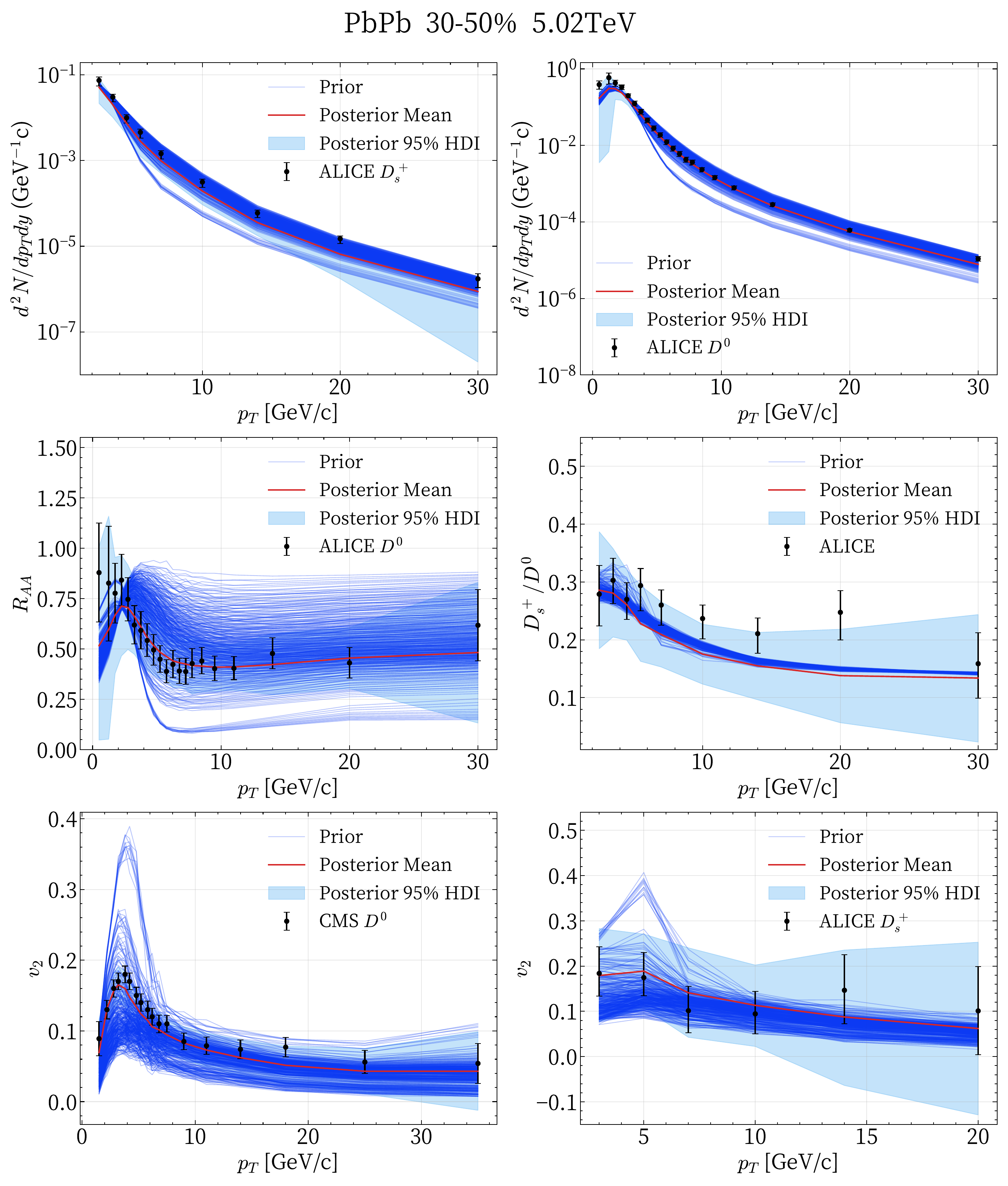}
\caption{30-50\% Pb-Pb collisions, $\sqrt{s_{NN}}=5.02$ TeV. Top-left: $D_s^+$ spectrum~\cite{ALICE:2021kfc}; Top-right: $D^0$ spectrum~\cite{ALICE:2021rxa}; Middle-left: $D^0$ $R_{AA}$~\cite{ALICE:2021rxa}; Middle-right: $D_s^+/D^0$ ratio~\cite{ALICE:2021kfc}; Bottom-left: $D^0$ $v_2$~\cite{CMS:2020bnz}; Bottom-right: $D^+_s$ $v_2$~\cite{ALICE:2021kfc}}
\label{pbpb3050-data}
\end{subfigure}
\caption{Prior and Posterior predictions of observables vs. data from ALICE and CMS with 95\% credibility intervals.}
\label{fig:combined-pbpb}
\end{figure*}

Figure~\ref{pbpb010-data} and Figure~\ref{pbpb3050-data} show the prior predictions from the direct model calculations and the posterior predictions after Bayesian inference with 95\% credibility intervals for the observables, compared with experimental data from ALICE~\cite{ALICE:2021kfc, ALICE:2021rxa} and CMS~\cite{CMS:2020bnz}, for 0-10\% and 30-50\% Pb-Pb collisions, respectively. Both the prior and posterior predictions are in reasonable agreement with the experimental data.

\end{document}